\documentclass[twocolumn,showpacs,preprintnumbers,amsmath,amssymb]{revtex4}
\usepackage{amsmath,amsfonts,graphics,epsfig,color}
\usepackage{graphicx}
\usepackage{dcolumn}
\usepackage{bm,textcomp}
\usepackage{bbm}
\usepackage{amssymb}
\usepackage{graphics}

\begin{document}


\title{Matrix Product States and Quantum Phase Transitions}

\author{Khabat Heshami, Sadegh Raeisi}
\affiliation{Department of Physics, Sharif University of Technology, P. O. Box: 11155-9161, Tehran, Iran.}

\date{\today}

\begin{abstract}

We have developed a new approach based on matrix product representations of ground states to study Quantum Phase Transitions (QPT). As confirmation of the power of our approach we have analytically analyzed the XXZ spin-one chain with uniaxial single-ion-type anisotropy and our results compare favourably with previous numerical studies. In addition, our description lets to know which part of parameters space of the Hamiltonian is most likely to be exactly solvable.

\end{abstract}

\maketitle

The challenging problem of determining the ground state of a quantum spin system and its properties such as the energy and degeneracy, and the behavior of associated correlation functions and QPTs \cite{sachdev} developed several numerical and analytical approaches in many-body physics. Numerically, the density matrix renormalization group (DMRG) \cite{DMRG} and analytically finitely correlated states \cite{FCS}, also known as translational invariant matrix product (TIMP) states \cite{MPRep, CiracSO, Cirac09, VahidXYZ, VahidMG} as generalization of valence bond states \cite{AKLT} and several other methods provide a valuable framework to deal with quantum spin chains, although still there is lack of specific formalism for studying QPT points. 

Recently, in \cite{Ciracqpt} a method was proposed to study a new family of QPTs based on the MPS formalism. They derived the correlation functions in terms of the eigenvalues of the transfer matrix $E$, and have shown that the correlation length $\xi=\frac{1}{\log{\mid \frac{v_1}{v_2}\mid}}$ diverges when level crossing occurs between two largest eigenvalues of $E$, $v_1$ and $v_2$. This is an interesting kind of QPTs which can be found for MP states \cite{Ciracqpt, VahidLad, VahidS1}. However, it is clear that in such QPTs there is no change in the ground state energy from one side of the critical point into the other side, because the ground state energy is always zero, which is far from a standard expectation for physically relevant QPTs.

In this letter, we propose a method to characterize QPTs based on MP representation of the ground state and we support our general discussions by comparing our analytic results with the numerical results of a well-known model for spin one chains. We also propose a formalism to find an MP representation for ground state of a given Hamiltonian, and present a criterion to determine in which part of the parameter space of the Hamiltonian the ground state is not an MPS. Based on this description we show that subspaces accompanied by the higher degeneracy in the local ground state are the most likely to yield the ground state exactly. This gives a new idea to find exact ground states of several models, and, using other methods like spin waves to determine ground state properties near the exactly solvable subspace. 

We can construct the parent Hamiltonian for 
$|\Psi\rangle=\frac{1}{\sqrt{Z}}\sum_{i_1...i_N=1}^{d}{\text{Tr}(A_{i_1}...A_{i_N})|i_1...i_N\rangle}
$, by finding $h^{(k)}=\Sigma_{i=1}^{r}{a_i|v_i\rangle\langle v_i|}$, which introduces the interaction among $k$ particles in a translational invariant parent Hamiltonian $H=\Sigma_{i=1}^{N}{h_{i...i+k}}$, where the $|v_i\rangle$'s span the null space of the reduced density matrix $\rho^{(k)}$ of $k$ neighboring
sites. For $a_i\geq 0$, $H$ is positive semidefinite and $|\Psi\rangle$ is the ground state with zero eigenvalue. Constructing the parent Hamiltonian for an MPS is based on the fact that the interaction among the particles can be modeled in a finite-dimensional vector space (which comes from the definition of the finitely correlated states). Despite this, the constructed parent Hamiltonian is typically not similar to well-known models, for example the Heisenberg model. So, it is important to study a formalism to find MP representations for the ground states of arbitrary models, and to aim at understanding in which part of the parameters space of a given Hamiltonian an MP representation for the ground state is possible.

We begin with the Hamiltonian of a translationally invariant local interaction,
\begin{equation}
H=\sum_{i=1}^{N}{h_{i...i+k}\otimes \mathbbmss{1}_{\rm{rest}}}.
\end{equation}
It is possible to represent the local interaction in terms of its eigenvalues and eigenvectors by spectral decomposition, $
h^{(k)}=\sum_{i=1}^{d^k}{\alpha_i |e_i\rangle\langle e_i|},$
and also, to determine the smallest eigenvalue of $h^{(k)}$. Let us consider the given Hamiltonian, with periodic boundary condition, as the sum of a positive semidefinite Hamiltonian and a constant term:
\begin{eqnarray}\label{Ham}
H &=& \sum_{i=1}^{N}{(h_{i...i+k}-\alpha \mathbbm{1}_{i..i+k})\otimes \mathbbmss{1}_{\rm{rest}}}+N\alpha \mathbbm{1}_{\rm{tot}}\\ \nonumber
&=& \sum_{i=1}^{N}{(h^{\prime}_{i...i+k}\otimes \mathbbmss{1}_{\rm{rest}} )}+N\alpha \mathbbm{1}_{\rm{tot}}\\ \nonumber
&=& H^{\prime}+N\alpha \mathbbm{1}_{\rm{tot}},
\end{eqnarray}
where $\alpha$ is the smallest eigenvalue of the $h^{(k)}$ and $H^{\prime}$ is a positive semidefinite operator, because
\begin{equation}\label{hprime}
h^{\prime(k)}\equiv\Sigma_{i=1}^{d^k-g}{(\alpha_i-\alpha) |e_i\rangle\langle e_i|}
\end{equation}
where $g$ is degeneracy of $\alpha$ and $\alpha_i>\alpha$.

Now, we pose the question: ``Is there any set of matrices $\{A_i\}$ to construct the MP $|\Psi\rangle$ such that $\langle\Psi|H^{\prime}|\Psi\rangle=0$?'' That is, can one arrange things such that the reduced density matrix of $|\Psi\rangle$ for $k$ consecutive sites has $|e_i\rangle$s as the basis for its null space? The MPS formalism indicates that the $|e_i\rangle$s impose a set of constraints on the $A_i$s, so this enables us to look for $d$ matrices to represent MP ground state for $H^{\prime}$ with zero eigenvalue. As we now discuss,  thinking along these lines yields a criterion -- Theorem 13 of \cite{Cirac09}-- to know whether there is a TIMPS as ground state.

It is clear that there is a bound on the ground state energy $E_g$ of $H$ for $N$ particles, namely $E_g\geq N\alpha$. Assuming that there exists an MPS such that $\langle\Psi_{MPS}|H'|\Psi_{MPS}\rangle=0$, it can be shown that $|\Psi_{MPS}\rangle$ is eigenstate of $H$ with $E_g=N\alpha$. Hence, if there exists $\{A_i\}_{i=1}^{d}$ that satisfy the equations imposed by the $\{|e_i\rangle\}_{i=1}^{d^k-g}$, then the ground state energy will be $E_g=N\alpha$. The greater the degeneracy in the local ground state impose the smaller the number ($d^k - g$) of constraints on the $A_i$s. Intuitively, the more degenerate subspaces in the space of parameters $\{J\}$ are ones for which it is more probable to find an exact ground state \cite{giorirapid, langari}. For example, AKLT have solved the $H=\sum_{i=1}^{N}(S_i.S_{i+1})-\beta (S_i.S_{i+1})^2$ for $\beta=-\frac{1}{3}$, exactly. It can easily be shown that $\beta=-\frac{1}{3}$ is the crossing point of the non-degenerate local ground state of energy $-2-4\beta$ (for $\beta > -\frac{1}{3}$) with the triply degenerate local ground state of energy $-1-\beta$ (for $\beta < -\frac{1}{3}$). So, at the level crossing point ($\beta=-\frac{1}{3}$) the local ground state degeneracy is maximum ($g=4$), and it can be shown that the local Hamiltonian consists of the $3^2-4=5$ remaining vectors which equal the spin-2 projector $P_{i,i+1}^{(2)}=\sum_{m=-2}^{2}\mid 2, m\rangle \langle 2,m\mid$. Fortunately, because of the continuity of $E_g$ in terms of the parameters of the Hamiltonian, {\it i.e.} coupling constants $\{J\}$, $E_g-N\alpha$ approaches zero, smoothly:
\begin{equation}\label{lim}
\lim_{\{J\}\rightarrow \{J\}_{b}} {E_g-N\alpha}=0,
\end{equation}
where, $\{J\}_b$ identifies the boundary of the region in which the ground state has TIMP representation. This investigation can be also performed using the DMRG \cite{DMRG} to calculate $E_g$. These facts allows us to look for regions in the parameter space of $H$, where the ground state is a TIMPS.When the ground state energy $E_g$ is equal to the minimum possible energy $N\alpha$, the ground state minimizes energy locally. This means that $\mathrm{Tr}(h^{(k)}\rho^{(k)})=\alpha$ or $\mathrm{Tr}(h^{\prime(k)}\rho^{(k)})=0$, where $h^{(k)}$ is the local Hamiltonian for $k$ neighboring sites, $h^{\prime(k)}$ is the positive local Hamiltonian in (\ref{hprime}) and $\rho^{(k)}$ is the reduced density matrix for $k$ sites. In another language $\rm{tr}\left( h^{\prime(k)} \rho^{(k)}\right)=0$ means that correlations between $k$ particles can be specified by a finite dimensional vector space. The equality $E_g=N\alpha$ leads to $\mathrm{tr}(h^{\prime(k)}\rho^{(k)})=0$ which means that $h^{\prime(k)}$ can be written in terms of the null space of $\rho^{(k)}$ \cite{Cirac09}, {\it i.e.}, the eigenvectors of the local Hamiltonian specify the null space of the reduced density matrix for each $k$-site block. So we can expect that, in these cases, the global ground state can be determined by studying local ground states like the MPS.

Taking a step further on (\ref{hprime}) for a given $H$, we now introduce a method to study QPTs. Consider the decomposition of $h^{\prime}$. It is clear that the eigenvectors corresponding to the smallest eigenvalue were omitted. Each remaining eigenvector results in a constraint equation on the matrices $A_i$s. For example, a vector like $|e\rangle=(1,a,b,c)$, which comes from a two body Hamiltonian for spin one-half particles, imposes an equation like $A_{+}A_{+}+aA_{+}A_{-}+bA_{-}A_{+}+cA_{-}A_{-}=0$, where $A_{\pm}$ denote matrices related to $|\pm\rangle$ states of spin one-half objects.

Therefore, a level crossing in the smallest eigenvalues of $h$ leads to at least one change in the eigenvectors that describe $h^{\prime}$, and so a change in the set of equations governing the $A_i$s. It is possible to have more alternations when $\alpha$ is degenerate. Solving these equations for a specific $D$, which is the dimension of $A_i$s, yields different MP ground states for each individual region, because they were obtained from different sets of equations. As such we can potentially identify a QPT purely by examining the structure of the low energy states of the $h$.

This is only a rough idea to give insight into our proposal. It is important to relate these rough ideas to discontinuity in a physical observable or define an order parameter to determine QPTs by this description.

Consider a model which has at least one level crossing in its smallest eigenvalue in terms of the Hamiltonian parameters. Thus, there are two corresponding regions and in each region, one of eigenvalues of $h$ is the smallest. At the level crossing point (it can be a point, a line, etc.) both of them are the smallest eigenvalues. Without trying to solve the equations in both regions, using Eq.~(\ref{lim}), it is possible to find out whether the ground state is Translational Invariant MPS or not. The cases in which the ground state in different regions are TIMPS, looking for discontinuity in derivatives of the ground state energy enables us to study QPTs. For instance, if $\alpha$ and $\beta$ are the smallest eigenvalues of $h$ and if the numerical investigations show that $E_g=N\alpha$ and $E_g=N\beta$ in the two regions,  any discontinuity in derivatives of $E_g$ - or non-zero derivatives of $\alpha-\beta$ in terms of $\{J\}$ - implies that a QPT occurs in the cross point.

There is a possibility of finding an order parameter in order to characterize QPTs in another way. Let us start with the example of a spin one-half Hamiltonian with a two body interaction. Consider a model that possesses $|e_1\rangle=(1,0,0,0)$ as eigenvector corresponding to the smallest eigenvalue $\alpha$ in one region, but are higher energy vector for the other region. This means that in the first region $A_{+}A_{+}\neq 0$ and in the other region $A_{+}A_{+}=0$, because $\alpha$ is not the smallest eigenvalue there. So, due to $|\Psi\rangle$, in the second region, the coefficient of all terms with consecutive up ($|+\rangle$) states are zero. In this case $O=\mid ++\rangle \langle ++ \mid$ is a suitable local operator to define an order parameter, because it can be shown that the quantity $\lim_{N\rightarrow \infty} {\Sigma_{i=1}^{N}{<O_{i,i+1}>}}$ is non-zero in the first region and zero in the other one.

Lets keep considering two body Hamiltonians, which in practise are the most physically interesting. The equations resulting from the spectral decomposition of the local Hamiltonian are linear combinations of the products $A_iA_j$. Clearly, these equations are nonlinear in terms of the $A_i$'s, and the parameters in them. But, it is possible to easily linearize these equations by replacing $A_iA_j$ with matrix $M_{ij}$. Again, consider that a level crossing occurs for the smallest eigenvalue of a local Hamiltonian. Now there are two sets of equations corresponding to the two possible regions in the space of parameters. The equations governing the $M_{ij}$s can be solved, which leads to the relation between $M_{ij}$s in both regions (after linearizing the equations). There are always solutions to the $M_{ij}$'s, however this does not mean that there are $A_i$s which satisfy all equations, because in the regions where $E_g>N\alpha$ solutions are accompanied by a contradiction between $M_{ij}=A_iA_j$ and $M_{ij}$s.

Now, based on the above example and discussions, we see that it is possible to define an order parameter by studying solutions of linearized sets of equations in models that possess a level crossing in their smallest eigenvalue. In general,
\begin{equation}
\lim_{N\rightarrow \infty}{\sum_{i=1}^{N}\langle O_{\alpha \beta}^{(i,i+1)}\rangle}
\end{equation}
corresponds to an operator $O_{\alpha \beta}=|\alpha\beta\rangle\langle\alpha\beta|$ which can be an order parameter when the sets of equations imply that $\exists \alpha ,\beta \backepsilon \{1,...,d\}$ such that $M_{\alpha \beta}=M_{\beta \alpha}=0$ in one region and $M_{\alpha \beta}=B_{\alpha}B_{\beta}\neq 0$ in another region. Clearly the above order parameter is zero in the region where $M_{\alpha\beta \neq 0}$ and is non-zero in another region. In other words, there is no ordering like $\mid \alpha \beta \alpha ... \rangle$ in one region while there is such a pattern in another region, even when there is no TIMP ground state in one of the regions.

In summary, the above condition yields a method to detect a QPT. This condition, based on the formalism for finding MP representations of ground states, lets us study non-trivial QPTs in some models. In the next section we study a well-known model for spin one systems and compare our analytic results with numerical outcomes.

We now apply our previous discussions on a well-known model, and analytically derive the presence of a QPT which previously was only known to exist through numerically studies. Consider the following two body Hamiltonian as the parent Hamiltonian for a spin one chain, which has been studied numerically in \cite{prb2003}.
\begin{equation}
H=\sum_{i=1}^{N}{S_x^iS_x^{i+1}+S_y^iS_y^{i+1}+J_zS_z^iS_z^{i+1}+D(S_z^i)^2}
\end{equation}
A spectral decomposition of the local Hamiltonian easily yields eigenvalues and eigenvectors.
The eigenvalues are
\begin{equation}
\left\{
\begin{array}{l}
e_1=J_z+D\\ \nonumber
e_2=(-J_z+D+\sqrt{J_z^2-2J_zD+D^2+8})/2\\ \nonumber
e_3=(-J_z+D-\sqrt{J_z^2-2J_zD+D^2+8})/2.\\ \nonumber
e_4=-J_z+D\\ \nonumber
e_5=(D+\sqrt{D^2+4})/2\\ \nonumber
e_6=(D-\sqrt{D^2+4})/2 \nonumber
\end{array}
\right.
\end{equation}
A simple calculation shows that there is a level crossing for the smallest eigenvalue, which is between $e_1$ and $e_3$. This means that there are two regions in the space of parameters $\{J_{z},D\}$. The following equation shows the line that separates these regions, which can be derived from $e_1-e_3=0$
\begin{equation}\label{line}
J_{z}=(-D-\sqrt{D^2+4})/2.
\end{equation}
Let us denote regions $R_{1}:J_{z}>(-D-\sqrt{D^2+4})/2$ and $R_{2}:J_{z}<(-D-\sqrt{D^2+4})/2$. In $R_1$ the smallest eigenvalue is $e_3$
and the corresponding eigenvector is $(0,0,1,0, -e_2, 0,1,0,0)$, whereas in $R_2$ the smallest eigenvalue is $e_1$ with $(1,0,0,0,0,0,0,0,0)$ and $(0,0,0,0,0,0,0,0,1)$ as eigenvectors.

Now, it is possible to look for TIMP ground state in each region by solving the corresponding sets of equations for each region. Omitting the above eigenvectors and translating the remained ones leads to the following sets of equations:
\begin{equation}
R_{1}:
\begin{array}{l}
M_{11}=0\\ \nonumber
M_{\bar{1}\bar{1}}=0\\ \nonumber
M_{1\bar{1}}-e_3M_{00}+M_{\bar{1}1}=0\\ \nonumber
M_{1\bar{1}}-M_{\bar{1}1}=0\\ \nonumber
e_6M_{0\bar{1}}-M_{\bar{1}0}=0\\ \nonumber
e_5M_{10}+M_{01}=0\\ \nonumber
e_5M_{0\bar{1}}-M_{\bar{1}0}=0\\ \nonumber
e_6M_{10}+M_{01}=0 \nonumber
\end{array}
R_{2}:
\begin{array}{l}
M_{1\bar{1}}-e_3M_{00}+M_{\bar{1}1}=0\\ \nonumber
M_{1\bar{1}}-e_2M_{00}+M_{\bar{1}1}=0\\ \nonumber
M_{1\bar{1}}-M_{\bar{1}1}=0\\ \nonumber
e_6M_{0\bar{1}}-M_{\bar{1}0}=0\\ \nonumber
e_5M_{10}+M_{01}=0\\ \nonumber
e_5M_{0\bar{1}}-M_{\bar{1}0}=0\\ \nonumber
e_6M_{10}+M_{01}=0 \nonumber
\end{array}
\end{equation}

\begin{figure}[ht]
\begin{center}
\epsfig{file=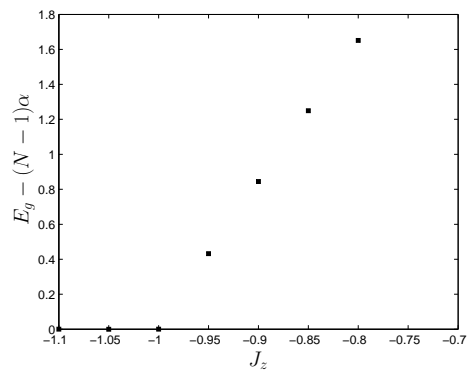,angle=0, width=0.7\linewidth}
\end{center}
\caption {$E_g-(N-1)\alpha$ in terms $J_z$ for an open chain for $N=30$ and $D=0$. For $J_z>-1$, in $R_1$, $\alpha=e_3$ and for $J_z<-1$, in $R_2$, $\alpha=e_1$. }
\label{figure}
\end{figure}
Numerical results for $N=30$ particles, shown in Fig. \ref{figure} indicate that in $R_2$ there is a TIMP representation for the ground state, whereas in $R_1$ DMRG shows that condition $Eg = (N-1)\alpha$ is violated. This means that the ground state is not of the TIMP form, so there is no solution for the equations governing the matrices to represent a TIMP ground state. Note that the numerical simulations using DMRG were performed for an open chain which clearly consists of an $N-1$ block so the $N$ in (\ref{lim}) must be replaced by $(N-1)$.

Since $E_g-(N-1)\alpha$ does not vanish in $R_1$, the ground state is a TIMPS only in $R_2$, and there is no solution with finite dimensional $A_i$s in $R_1$. Solving the set of equations governing the $A_i$s in $R_2$ shows that only $M_{11}=A_1A_1$ and $M_{{\bar 1}{\bar 1}}=A_{{\bar 1}}A_{{\bar 1}}$ are non-zero. This means that $R_2$ is the ferromagnetic phase of this model. It is clear that $M_{11}$ and $M_{{\bar 1}{\bar 1}}$ are zero in $R_2$, so local operators $O_{11}=|11\rangle\langle11|$ and $O_{{\bar 1}{\bar 1}}=|{\bar 1}{\bar 1}\rangle\langle{\bar 1}{\bar 1}|$  - and any linear combination of them - are suitable choices to show that a QPT occurs in the boundary (\ref{line}) between $R_1$ and $R_2$. In this case one can select the more familiar operator $S_z^{i}$.
\begin{figure}[t]
\begin{center}
\epsfig{file=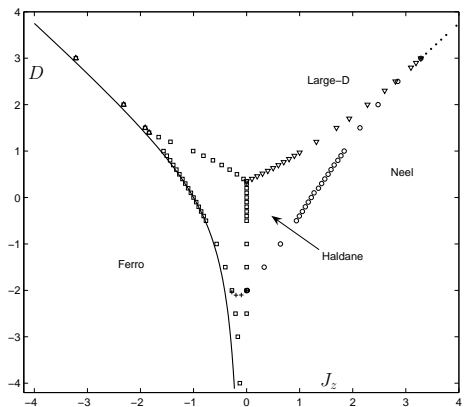,angle=0, width=0.7\linewidth}
\end{center}
\caption {The solid line is Eq.~(\ref{line}) and is the result of our consideration based on changes in the structure of equations governing matrix product representations of the ground state. The other parts of the figure come from numerical studies performed in \cite{prb2003}.}
\label{qpt}
\end{figure}
Hence, the line (\ref{line}) demonstrates a boundary between a ferromagnetic phase and other (generally anti-ferromagnetic) phases, of this model. Figure~\ref{qpt} shows our analytical result in comparison to numerical studies using exact diagonalization in \cite{prb2003} which is a good evidence for confidentiality of our results.

We have proposed a method for studying spin chains with local interactions which in certain circumstances allows one to identify regions in parameter space for which  there is no TIMP representation of the ground state. In addition, we have shown that in the space of parameters of the local Hamiltonian, subspaces where the local ground state has the highest degree of degeneracy are the most likely to be exactly solvable which can be guiding in further studies. Furthermore, it is possible to use this description for identifying and studying quantum phase transitions. In contrast to a family of QPTs that have been studied in \cite{Ciracqpt}, QPTs that can be considered through this description are accompanied by a change in energy. 

We can use this formalism to study one dimensional spin systems and to benefit from the simplicity of the Matrix Product representation to calculate correlation functions, correlation length or investigate entanglement in these systems. In addition, the possibility of finding an approximate solution for the set of equations, {\it i.e.}, a TIMPS which approximately describes the ground state for example by $A_i$s satisfying ${\rm tr}(\sum_{i,j=1..d}{c_{ij}A_iA_j})=\epsilon$ for a small finite $\epsilon$ rather than $\epsilon=0$ can be studied.

Finally we mention that it is possible to turn things around and impose level crossings of the smallest eigenvalue to construct and engineer new models with a TIMP ground state accompanied by determined Quantum Phase Transition, and study behavior of correlations and entanglement near critical points.

\textit{Acknowledgement--} We acknowledge Terry Rudolph, Tobias Osborne, Vahid Karimipour, Abdollah Langari, Sima Baghbanzadeh and Laleh Memarzadeh for their valuable discussions and comments, authors also thank Wei Chen for the data of their numerical study.


\begin{thebibliography}{99}
\bibitem{sachdev} S. Sachdev, {\it Quantum Phase Transitions}
(Cambridge Univ. Press, Cambridge, 1999).
\bibitem{DMRG} S. R. White, Phys. Rev. Lett {\bf 69}, 2863 (1992).
\bibitem{FCS} M. Fannes, B. Nachtergaele and R. W. Werner, Comm.
Math. Phys. {\bf 144}, 443 (1992).
\bibitem{MPRep} D. Perez-Garcia, F. Verstraete, M. M. Wolf and J. I.
Cirac, Quantum Inf. Comput. {\bf 7}, 401 (2007).
\bibitem{CiracSO} D. Perez-Garcia, M. M. Wolf, M. Sanz, F. Verstraete and J. I. Cirac, Phys. Rev. Lett. {\bf 100}, 167202 (2008).
\bibitem{VahidMG} V. Karimipour and L. Memarzadeh, Phys. Rev. B {\bf 77},
094416 (2008).
\bibitem{VahidXYZ} M. Asoudeh, V. Karimipour and A. Sadrolashrafi,
Phys. Rev. A {\bf 76}, 012320 (2007).
\bibitem{Cirac09} M. Sanz, M. M. Wolf, D. Perez-Garcia, and J. I. Cirac, arxiv:condmat/0901.2223v1 (2009).
\bibitem{AKLT} I. Affleck, T. Kennedy, E.H. Lieb, H. Tasaki,
Commun. Math. Phys. {\bf 115}, 477 (1988); I. Affleck, T. Kennedy, E.H. Lieb, H. Tasaki, Phys. Rev. Lett. {\bf
59}, 799 (1987).
\bibitem{Ciracqpt} M. M. Wolf, G. Ortiz, F. Verstraete and J. I. Cirac,
Phys. Rev. Lett. {\bf 97}, 110403 (2006).
\bibitem{VahidLad} M. Asoudeh, V. Karimipour and A. Sadrolashrafi,
Phys. Rev. B {\bf 75}, 224427 (2007).
\bibitem{VahidS1} S. Alipour, V. Karimipour and L. Memarzadeh, Phys. Rev. A {\bf 75}, 052322 (2007).
\bibitem{giorirapid} G. L. Giorgi, Phys. Rev. B {\bf 79}, 060405(R) (2009).
\bibitem{langari} M. Rezai, A. Langari and J. Abouie, arxiv: condmat/0904.3843v1 (2009).
\bibitem{prb2003} W. Chen, K. Hida, B. C. Sanctuary, Phys. Rev. B {\bf 67}, 104401 (2003).
\end{thebibliography}
\end{document}